\documentclass[prl,letterpaper,twocolumn,showpacs,
preprintnumbers,citeautoscript,amsmath,amssymb]{revtex4}

\usepackage{graphicx}
\usepackage{dcolumn}

\begin{document}
\title{
Exceptionally large room-temperature ferroelectric polarization in the novel PbNiO$_3$ multiferroic oxide.}
\author{X.F. Hao$^{1}$}
\author{A. Stroppa$^{2}$}
\author{S. Picozzi$^{2}$}
\author{A. Filippetti$^{3}$}
\author{C. Franchini$^{1}$}
\email[Corresponding author:]{cesare.franchini@univie.ac.at}
\affiliation{$^{1}$Faculty of Physics, University of Vienna and Center for
Computational Materials Science, A-1090 Wien, Austria}
\affiliation{$^{2}$CNR-SPIN L'Aquila, Via Vetoio 10, I-67100 L'Aquila, Italy}
\affiliation{$^{3}$CNR-IOM, UOS Cagliari and Dipartimento di Fisica, Universit\`a di Cagliari, Monserrato (CA), Italy}

\date{\today}

\begin{abstract}
We present a study based on several advanced First-Principles methods, of the recently synthesized PbNiO$_{3}$ [J. Am. Chem. Soc \textbf{133}, 16920 (2011)], a rhombohedral antiferromagnetic insulator which crystallizes in the highly distorted $R3c$ crystal structure. We find this compound electrically polarized, with a very large electric polarization of $\sim$ 100 $\mu$C/cm$^2$, thus even exceeding the polarization of well-known BiFeO$_{3}$. PbNiO$_{3}$ is a proper ferroelectric, with polarization driven by large Pb-O polar displacements along the [111] direction. Contrarily to naive expectations, a definite ionic charge of 4+ for Pb ion can not be assigned, and in fact the large Pb 6$s$-O 2$p$ hybridization drives the ferroelectric distortion through a lone-pair mechanism similar to that of other Pb- and Bi-based multiferroics.
\end{abstract}


\maketitle

\textit{Introduction}
Multiferroics are materials in which different ferroic orders 
such as ferromagnetism, ferroelectricity and/or ferroelasticity 
may coexist in one single phase~\cite{Khomskii}. In the last
few years, there has been a tremendous boom of interest in these 
materials, due to the potential applications 
in memory devices or in novel type of magnetic switching in magnetoelectric 
multiferroics, based on the cross-coupling between ferroelectric and magnetic 
channels~\cite{Tokura,Cheong,Mostovoy}.
Furthermore, these materials offer a rich and 
fascinating playground for the complex physical mechanisms  
underlying the processes involved in the observed properties.\cite{Bersuker}
It is obvious that  the search and 
the prediction by material design of
new multiferroics is of  great importance 
for both fundamental physics and technological 
applications~\cite{Rabe,Fennie}. 

\begin{figure}
\includegraphics[width=1.0\columnwidth,scale=1.0]{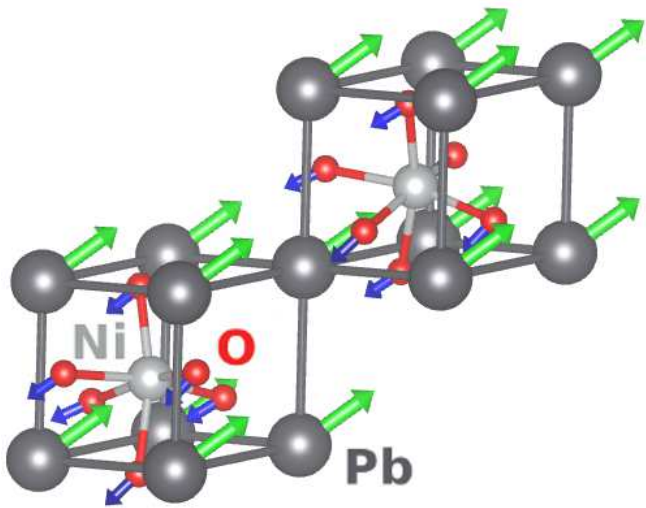}
\caption{\label{FIG:1}
(Color online) Schematic view of the LiNbO$_3$-type 
PbNiO$_3$ within space group {\em R3c} structure built
 up from two cubic perovskite unit cells,
 the black (large), gray (medium) and red
(small) spheres denotes Pb, Ni and O atoms, 
respectively. With respect to the centrosymmetric ({\em R\={3}c}) paraelectric phase (see text) the cations are displaced along the 
[111] axis relative to the anions (the arrows represent the displacement vectors), and the oxygen 
octahedra slightly rotate with alternating sense around
the [111] direction~\cite{Inaguma}.  
The crystal structure was drawn using the program VESTA\cite{vesta}.
}
\end{figure}

Relatively few multiferroics have been identified so far.~\cite{Hill}
Without doubt, the most studied and well characterized multiferroic is BiFeO$_{3}$~\cite{BFO}.
It crystallizes in the polar space group {\em R3c} (No. 161, point group {\em C}$_3$$_v$), 
and is predicted to have a G-type antiferromagnetic (AFM) alignment of the Fe spins~\cite{Ghosez,Rabe}.
The {\em R3c} symmetry corresponds to the so-called LiNbO$_3$-type structure,        
which can be viewed as a highly distorted double perovskite with rhombohedral symmetry. 
The primitive unit cell contains two formula units (10-atoms), arising
from counterrotations of neighboring O octahedra about the [111] axis (see Fig.\ref{FIG:1}). 
The crystal symmetry allows the presence of a spontanous polarization along the 
[111] direction, which arise from the relative displacement of the Bi    
sublattices with respect to the FeO$_{6}$ octahedra cages along [111].
The origin of the large spontaneous polarization of Bi$^{+3}$Fe$^{+3}$O$^{-2}$$_3$,
$\sim$ 90 $\mu$C/cm$^2$~\cite{BFO}, has been explained by first principles
calculations within the density functional theory plus Hubbard-U approach (DFT+U)\cite{Anisimov}
and the "modern theory of polarization" (MTP)~\cite{Resta,Vanderbilt}, in terms of the 
lone-pair electrons present at the Bi sites which are ultimately responsible for the 
large displacement along the [111] direction~\cite{Rabe}.

Inspired by the recent 
reports of Inaguma {\em et al.} on the synthesis of  a new antiferromagnetically ordered compound with LiNbO$_3$-type structure, such as  PbNiO$_3$\cite{Inaguma,Inaguma-2}, we explore here  the possibility of multiferroic behavior in PbNiO$_3$.
We first summarize the experimental findings.
Inaguma and coworkers  synthesized two high-pressure polymorphs of PbNiO$_3$ with a
(A) perovskite-type structure and (B) the LiNbO$_3$-type structure~\cite{Inaguma,Inaguma-2};
the latter (hereafter called L-PbNiO$_{3}$) is thermodynamically 
more stable than the perovskite-type one at ambient pressure.
With respect to the orthorhombic structure, in the acentric rhombohedral  LiNbO$_3$-type ({\em R3c}) structure (Fig.~\ref{FIG:1}) Pb and O atoms are displaced against each other along the threefold [111] axis 
leading to a large distortion of the PbO$_6$ and NiO$_6$ octahedra (see Fig.\ref{FIG:1}). 
The Pb atom is coordinated by six oxygens at 2.10 and 2.25 \AA, while the Ni--O bond distance splits 
into two subgroups (2.07 and 2.11 \AA). Magnetic susceptibility and resistivity measurements 
indicate that L-PbNiO$_3$ undergoes an AFM transition at $T_\mathrm{N}$=205$^{\circ}$K,
and exhibits semiconducting behavior. The AFM ordering in the acentric crystal structure  suggests 
possible multiferroic behavior.

In the following we show that L-PbNiO$_{3}$ is a \textit{new} room-temperature \textit{multiferroic}, with an \textit{exceptionally large} polarization (${\bf P}$) of about 100 $\mu$C/cm$^{2}$, which is the highest polarization ever predicted for any bulk material so far, and $\approx$ 10\% larger than that of BiFeO$_{3}$~\cite{BFO}. 

\textit{Computational details}
We performed DFT based calculations using the Vienna {\em Ab initio} Simulation Package 
(VASP)~\cite{Kresse_vasp} based on the projector-augmented-wave method\cite{PAW1,PAW2},
withing the Perdew, Burke and Ernzerhof (PBE) parameterization scheme\cite{Perdew} for the 
generalized gradient approximation (GGA). 
To overcome the deficiencies of standard exchange-correlation approximations for localized
Ni {\em d} states, we made use of three beyond-DFT approaches: 
(i) Dudarev's GGA+U~\cite{Anisimov,Dudarev}, using U=4.6 eV in accordance with constrained DFT
calculation of Ref.\onlinecite{Cococcioni};
(ii) The renowned Heyd-Scuseria-Ernzerhof (HSE) screened hybrid functional~\cite{HSE},
which has been shown to give an excellent account of materials
properties for magnetic multiferroics\cite{Stroppa}.
(iii) The recently introduced variational pseudo self-interaction-correction method VPSIC\cite{vpsic} 
(adopting the HSE optimized structure), implemented within
plane-wave basis set and ultrasoft pseudopotential scheme in the PWSIC code.
The cutoff energy was set as 600 eV and a 8$\times$8$\times$8 Monkhorst-Pack
grid of {\em k} points was used. In HSE, the fraction (1/4) of Fock exchange was sampled using the 
twofold reduced {\bf k}-point grid, to reduce the computational load.
The lattice parameters and atomic positions were relaxed (at GGA+U and HSE level) until the total energy changed by less 
than 10$^-$$^5$ eV per unit cell and the residual force was smaller than 0.01 eV/\AA.

%%%%%%%%%%%%%%%%%%%%%%%%%%%%%%%%%%%%%%%%%%%%%%%%%%%%
\begin{table}
\caption{Structural parameters, magnetic moment and electronic band gap for AFM-G configuration 
of L-PbNiO$_3$ within  GGA, GGA+U and HSE method, compared with the available
experimental data. {\em a}, {\em c} are the lattice parameters in the hexagonal setting, whereas  
{\em x}, {\em y}, and {\em z} are the internal atomic positions of Pb and O. Ni ions sit in the 
($6a$) (0,0,0) positions. Bond lengths of Pb--O and Ni--O are also reported.
$\widehat{{\rm Ni}-{\rm O}-{\rm Ni}}$ ($^{\circ}$) is the Ni--O--Ni angle.
{\em m} indicates the magnetic moment of Ni.}
\vspace{0.3cm}
\begin{ruledtabular}
\begin{tabular}{lccccc}
                                                 & GGA     & GGA+U   & HSE     & Expt.   \\
$a$ (\AA)                                        & 5.442   & 5.430   & 5.359   & 5.363   \\
$c$ (\AA)                                        & 13.737  & 14.353  & 14.209  & 14.090  \\
${z_{\rm Pb}}$                                   & 0.2226  & 0.2885  & 0.2898  & 0.2864  \\
${x_{\rm O}}$                                    & 0.0605  & 0.0532  & 0.0468  & 0.0487  \\
${y_{\rm O}}$                                    & 0.2598  & 0.3543  & 0.3579  & 0.3657  \\
${z_{\rm O}}$                                    & 0.0931  & 0.0702  & 0.0687  & 0.0668  \\
Ni-O (\AA)                                       & 2.051   & 2.128   & 2.119   & 2.109   \\
                                                 & 1.905   & 2.060   & 2.053   & 2.071   \\
Pb-O (\AA)                                       & 2.395   & 2.310   & 2.258   & 2.246   \\
                                                 & 2.238   & 2.174   & 2.121   & 2.104   \\
$\widehat{{\rm Ni}-{\rm O}-{\rm Ni}}$ ($^{\circ}$)& 147.8  & 140.6   & 138.1   & 136.7   \\ 
$m_{\rm Ni}$ ($\mu_{\rm B}$)                     & 0.97    & 1.67    & 1.69    & ---     \\
gap (eV)                                         & Metallic& 0.37    & 1.18    & ---     \\
\end{tabular}
\end{ruledtabular}
\label{tab:1}
\end{table}
%%%%%%%%%%%%%%%%%%%%%%%%%%%%%%%%%%%%%%%%%%%%%%%%%%%%
\textit{Magnetic, structural and electronic properties}
The G-type antiferromagnetic configuration (AFM-G), in which each Ni ion is surrounded 
by six nearest-neighbors Ni with opposite spin direction, is the most favourable 
spin configuration with respect to ferromagnetic (by 50 meV/f.u.), 
AFM type-A (by 33 meV/f.u.) and AFM type-C (by 17 meV/f.u.) orderings.

The resulting calculated magnetic Ni magnetic moment ($m_{\rm Ni}$) is about 1.7 $\mu_{\rm B}$ at both GGA+U and HSE level. The optimized GGA, GGA+U and HSE structural parameters for the AFM-G phase 
are listed in Table \ref{tab:1}. As expected, GGA does not reproduce well the experimental values, whereas 
HSE and, to a lesser extent GGA+U, deliver numbers in good agreement with experiment,
with relative errors of 1.5-2 \% (GGA+U) and $<$ 1.0\% (HSE).

The drawbacks of conventional GGA are more dramatic for the electronic properties.
The GGA scheme finds a metallic character, in clear disagreement with experiment~\cite{Inaguma}.
Conversely, the inclusion of either the on-site U or the Fock exchange cures this limitation and correctly predicts an insulating state with an energy gap of $\sim$ 0.4 eV (GGA+U) and $\sim$ 1.2 eV (HSE), as inferred from the density of states (DOS) shown in Fig.\ref{FIG:3} (the VPSIC gap is 0.9 eV, see Supplementary Materials (SM)). The gap opens between occupied Ni $d$ and O $p$ states, and the lowest unoccupied electronic states that are made up by a mixture of Pb $s$ and O $s$ states.

Spin polarization does not affect Pb and O DOS, whereas the Ni atoms display a magnetic moment of 
1.69 $\mu_{\rm B}$. In hexagonal coordinates (with $z$ parallel to the [111] direction) the Ni $d$ states are distributed into two doublets, ($d_{xy}$, $d_{x^2-y^2}$) and ($d_{xz}$, $d_{yz}$), which contribute to the net magnetization by 0.65 and 1.65 $\mu_{\rm B}$, respectively, and a spin-degenerate $d_{z^2}$ singlet (see SM for an $lm$-projected DOS). The broad Ni 3$d$ valence manifold is strongly hybridized with the O 2$p$ states, and lies in the energy range between -6 eV and the Fermi level. 

The most important feature of the DOS is at the bottom of the valence region, between -9 eV and -6 eV:
a large region dominated by highly-hybridized Pb 6$s$ and O 2$p$ states. The presence of a substantial amount of charge (about 1 electron, see SM) within the Pb 6$s$ orbital is at odds with the nominal Pb$^{4+}$ valence originally assumed in the interpretation of X-ray photoemission spectroscopy (XPS) results\cite{Inaguma,Inaguma-2}. In realistic calculations, substantial deviation between the static charges and the nonimal ionic charges (due to large hybridization effects) is the rule rather than the exception, thus it certainly does not come as a surprise. In this case, however, this deviation is worthy to be emphasized, 
since, as we will show below, it is strictly connected to the ferroelectric instability.
 To highlight the role of Pb 6$s$-O 2$p$ hybridization, in Fig.\ref{FIG:3} we  show the DOS for the centrosymmetric, non-polar reference structure {\em R\={3}c} as well as the polar structure: the major difference is indeed the 2 eV downshift and the broadening (more than a factor-2 bandwith expansion) of Pb 6$s$-O 2$p$ spectral weight occuring along with the centrosymmetric-to-ferroelectric transformation. This so-called stereochemical activity of the A-site cation \cite{Hill99, Seshadri} is quite consistent with what occurs in Bi-based (BiFeO$_3$\cite{Rabe}, BiMnO$_3$\cite{Hill99,Seshadri}) and Pb-based (PbTiO$_3$\cite{Cohen}) compounds, usually labelled as lone-pair ferroelectrics. (see SM for a more detailed description of the bonding picture in PbNiO$_3$ and a one-to-one comparison with the BiFeO$_3$ case). This is also reflected in the structural distortions which {\em predominantly} involve counterdisplacements of Pb and O atoms along the [111] direction (see Fig.\ref{FIG:1}).  

In the following we explore the ferroelectric properties of L-PbNiO$_3$ by evaluating the spontaneous 
macroscopic polarization {\bf P}.~\cite{Rabe}. 

\begin{figure}
\includegraphics[clip=, width=0.9\columnwidth,scale=1.0]{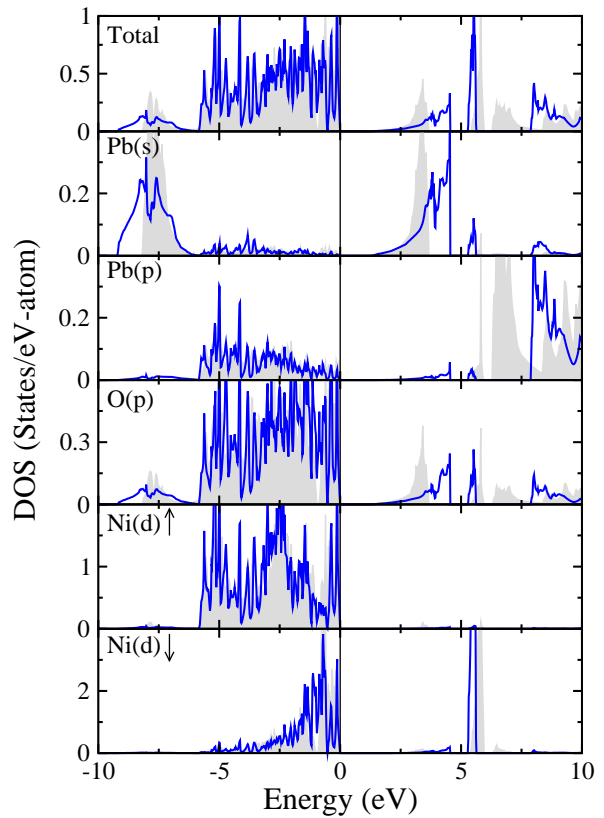}              
\caption{ \label{FIG:3}
(Color online) HSE calculated total and site-decomposed density of states for the 
AFM-G type PbNiO$_3$ in the acentric {\em R3c} (full lines) and centric {\em R\={3}c} (shadow)
phases. The vertical line denotes to the Fermi level.} 
\end{figure}

\textit{Analysis of ferroelectric polarization}

By pseudosymmetry analysis\cite{bilbao} we determine a parent centrosymmetric structure ({\em R\={3}c}) with minimal supergroup symmetries, from which the observed non-centrosymmetric structure can be reached through a continuous structural distortion. By definition, the polarization is zero (modulo a polarization quantum\cite{Resta,Vanderbilt}) in this paraelectric (PE) phase ($\lambda=0$), thus it represents a viable reference state for the evaluation of the polarization in the polar structure ($\lambda=1$). Here $\lambda=1$ is the amplitude of the polar distortion which progressively transforms the {\em R\={3}c} phase into the {\em R3c} one. For consistency, we assume for the {\em R\={3}c} phase the same volume and rhombohedral angle of the polar structure.

\begin{figure}
\includegraphics[clip=, width=1\columnwidth,scale=1.0]{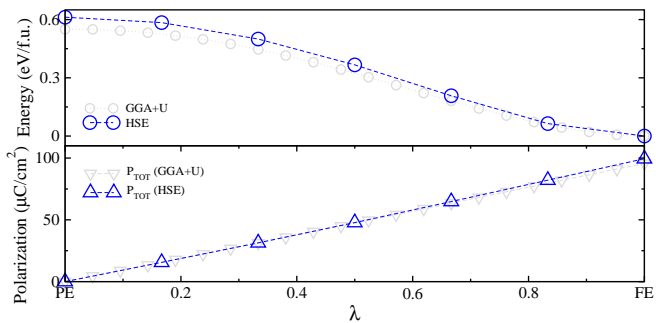}
\caption{\label{FIG:2}
(Color online) Total energy profile {\em E}, total polarization ${\bf P}_\mathrm{tot}$ as a function of the
polar distortion $\lambda$, from the $\lambda$=0 PE phase to the $\lambda$=1 ferroelectric (FE) one.
Both HSE (dashed lines, dark) and GGA+U (dotted lines, gray) data are displayed.}
\end{figure}

In the PE phase, Pb and O atoms lie in (111) PbO$_3$ planes. Symmetry mode analysis shows that the $\lambda=[0,1]$ transformation (showed by arrows in Fig.1) is almost exclusively composed by in-phase [111]-parallel relative shift of Pb and O atoms (i.e. a nearly-exclusive $\Gamma_{2-}$-mode transformation), with some very residual O-rotation contribution. The Pb atoms are displaced from their positions by 0.56 \AA~and the O atoms by 0.23~\AA, while Ni atoms are almost unchanged.
This transformation produces an alternate shrinking and dilatation of Pb-O distances along [111] of about 0.8 \AA, whereas Ni (111) planes are left off-centered by 0.17 \AA~with respect to their adjacent PbO$_3$ planes. 
The total polarization {\bf P} can be split as the sum of ionic P$_{ion}$ and electronic contribution 
P$_{ele}$: the former is the dipole of the ion-core charges; the latter is obtained by Berry phase approach within the MTP~\cite{Vanderbilt,Resta}. HSE and VPSIC describe both polar and nonpolar structures as insulating, whereas within GGA+U a large U=7.6 eV is required to open a gap in the PE phase. The HSE-calculated DOS shown in Fig.\ref{FIG:3} indicates that the value of the gap is essentially the same for both acentric {\em R3c} and nonpolar {\em R\={3}c} phases.
Fig.~\ref{FIG:2} displays total energy and polarization ${\bf P}_\mathrm{tot}$
as a function of the polar distortion $\lambda$ at both HSE and PBE+U level. Both methods deliver the same outcome. As expected the polar structure is more stable than the non-polar one by 0.6 eV/f.u (0.7 eV/f.u. according to VPSIC). Both ${\bf P}_\mathrm{ion}$ and ${\bf P}_\mathrm{ele}$ grow monotonically with the polar distortion, giving rise to a total polarization ${\bf P}_\mathrm{tot}$ of $\sim$ 100 $\mu$C/cm$^2$ (98.5 $\mu$C/cm$^2$ according to VPSIC), thus $\approx$ 10\% larger than that of BiFeO$_3$~\cite{Rabe,Stroppa} and significantly enhanched with respect to that of LiNbO$_3$ (80 $\mu$C/cm$^2$)\cite{Ghosez} and ZnSnO$_3$ (57 $\mu$C/cm$^2$)\cite{Nakayama}.

We have also evaluated the polarization using the approximated expression ${\bf P}_\mathrm{tot}$ = $\sum_i$ $\Delta R_i$ Z$^*_i$, where $\Delta R_i$ are atomic displacements (from centrosymmetric to polar structure structure) and Z$^*_i$ are Born effective charges (BEC) calculated via linear-response formalism in GGA+U. We obtain a value (93.8 $\mu$C/cm$^2$) substantially similar to the evaluation based on the exact Berry phase formula. 
 We found a nearly isotropic BEC tensor and dynamical charges similar in both PE and FE phases (see Tab.\ref{tab:born}). For Pb we obtain an anomalous BEC value of 4.4 (as compared to the static charge, $\approx$ 2.15), identical to the Bi BEC calculated in BiFeO$_3$, coherently with the role of this ion in guiding the ferroelectric transition\cite{BFO}.

\textit{Summary}
Using several beyond-LDA methods (LDA+U, HSE, and VPSIC) and the MTP approach for the determination of polarization properties, we report the presence of a large spontaneous electric polarization $\sim$ 100 $\mu$C/cm$^2$ in rhombohedral PbNiO$_3$ associated to the structural transformation from the centrosymmetric {\em R\={3}c} to the polar {\em R3c} symmetry. The microscopic analysis shows that this proper ferroelectric instability is dominated by huge in-phase (i.e. $\Gamma$-mode) relative displacements of Pb and on-top O atoms along the [111] direction, in turn associated to large Pb $s$-O $p$ band rehybridization at the bottom of valence band manifold. This ferroelectric mechanism can be then assimilated to other Bi-based and Pb-based ferroelectrics, whereas the naive interpretation of Pb as an inactive $+4$-charged ion would be totally misleading. This material may be prototype of a new class of Ni-based rhombohedral multiferroics, which take advantage of the stable rhombohedral symmetry to develop large ferroelectric displacements along the [111] axis, and still mantain strong magnetic coupling in the Ni sublattice.

%%%%%%%%%%%%%%%%%%%%%%%%%%%%%%%%%%%%%%%%%%%%%%%%%%%%
\begin{table}
\caption{Born effective charges for Pb and Ni in PbNiO$_3$ (with ferroelectric and paraelectric phases)
computed via linear-response formalism within the GGA+U method.}
\vspace{0.3cm}
\begin{ruledtabular}
\begin{tabular}{lccc}
                    &  Pb  &  Ni  & O         \\
{\em R{3}c} (FE)     & 4.40 & 1.97 & -2.12     \\
{\em R\={3}c} (PE)  & 4.41 & 1.97 & -2.12
\end{tabular}
\end{ruledtabular}
\label{tab:born}
\end{table}
%%%%%%%%%%%%%%%%%%%%%%%%%%%%%%%%%%%%%%%%%%%%%%%%%%%%

%%%%%%%%%%%%%%%%%%%%%%%%%%% Acknowledgment%%%%%%%%%%%%%%%%%%%%%%%%%%%%%%%%%
%--------------------------
\section*{Acknowledgements}
%--------------------------
Support by European Community (FP7 EU-INDIA grant ATHENA and ERC Starting Grant 
no.203523 BISMUTH ) is gratefully acknowledged.
X.H. thanks Claude Ederer and  Chung-Yuan Ren for their helpful advices and comments.
A.S. thanks J.M. Perez-Mato for useful discussions.
The calculations have been performed on the Vienna Scientific Cluster (VSC) 
and, partially, in CASPUR Supercomputing Center in Rome.
%%%%%%%%%%%%%%%%%%%%%%%%%%%%%%%%%%%%%%%%%%%%%%%%%%%%%%%%%%%%%%%%%%%%%%%%%%%%

\end{document}